# Tissue fibrosis: a principal proof for the central role of Misrepair in aging

Jicun Wang-Michelitsch[1]*, Thomas M Michelitsch[2]

[1]Department of Medicine, Addenbrooke's Hospital, University of Cambridge, UK (Work address until 2007)

[2] Institut Jean le Rond d'Alembert (Paris 6), CNRS UMR 7190 Paris, France

**Abstract**

Tissue fibrosis is the phenomenon that a tissue has progressive deposition of collagen fibers with age. Tissue fibrosis is associated with aging of most of our organs, and it is the main pathology in arteriosclerosis, chronic bronchitis/emphysema, and benign prostatic hyperplasia. The causes and characteristics of fibrosis are analyzed with Misrepair mechanism, a mechanism proposed in Misrepair-accumulation aging theory. Tissue fibrosis is known to be a result of repairs of tissue by collagen fibers. A repair with collagen fibers is a manner of "Misrepair". The collagen fibers are used for replacing dead cells or disrupted extracellular matrixes (ECMs) including elastic fibers, myofibers, and basement membrane. The progressive tissue fibrosis with age manifests the essential role of Misrepair in aging, because it reveals three facts: **A**. a process of Misrepair exists; **B**. Misrepairs are unavoidable; and **C**. Misrepairs accumulate. As a result of accumulation of Misrepairs of tissue with collagen fibers, tissue fibrosis is focalized and self-accelerating, appearing as growing of spots of hyaline degeneration. Fibrosis results in stiffness or atrophy of an organ and progressive failure of the organ. In arteriosclerosis, the deposition of collagen fibers in arterial wall is for replacing disrupted elastic fibers or myofibers, however results in hardness of the wall. In benign prostatic hyperplasia, the deposition of collagen fibers in prostate is for replacing broken myofibers, however results in stiffness and weakness of the muscular tissue and deposition of prostatic fluid in gland tubes. Wrinkle formation is part of skin fibrosis, and it may be a result of accumulation of collagen fibers of different lengths. Senile hair-loss and hair-whitening are probably consequence of dermal fibrosis. In conclusion, tissue fibrosis is a result of accumulation of Misrepairs of tissue with collagen fibers, and the phenomenon of fibrosis is a powerful proof for the central role of Misrepair in aging.

**Keywords**

Tissue fibrosis is the phenomenon that a tissue has progressive deposition of excessive collagen fibers with age. Tissue fibrosis takes place in most of our organs, and it is associated with development of many diseases and aging changes. Fibrosis is the main cause for the diseases of essential arterial hypertension (arteriosclerosis), senile chronic bronchitis/emphysema, and benign prostatic hyperplasia. It is reported that more than 45% of world-wide deaths are related to fibrosis of organs. Tissue fibrosis is a typical aging change; however none of traditional aging theories is able to give a satisfactory interpretation on it. In the present paper, we will discuss the causes and the characteristics of tissue fibrosis with Misrepair mechanism, a mechanism proposed in Misrepair-accumulation aging theory (Wang, 2009). Our discussion tackles the following issues:

I.   Fibrosis-related diseases

II.  A novel aging theory: Misrepair-accumulation theory

III. Tissue fibrosis: a result of accumulation of Misrepairs of tissue with collagen fibers

   3.1  Collagen fibers in replacing dead cells and broken basement membrane
   3.2  Collagen fibers in replacing disrupted elastic fibers and myofibers
   3.3  Collagen fibers in isolating un-degradable substances in a tissue
   3.4  Accumulation of repairing collagen fibers: focalized and self-accelerating

IV.  Interpretation of fibrosis-related aging changes

   4.1  Arteriosclerosis develops as a result of accumulation of collagen fibers in arterial wall
   4.2  Benign prostatic hyperplasia develops as a consequence of muscular fibrosis in prostate
   4.3  Wrinkle formation is a result of accumulation of collagen fibers of different lengths
   4.4  Senile hair-loss and hair-whitening are consequences of dermal fibrosis
   4.5  Senile atrophy of the brain is due to "fibrosis" of nerve tissue?

V.   Conclusions

## I.   Fibrosis-related diseases

Tissue fibrosis has been observed in most aged organs in human being. There are two manners of depositions of collagen fibers in a fibrotic tissue: **A**. without loss of parenchyma cells; and **B**. with loss of parenchyma cells. The first one takes place often in an "elastic" organ, such as artery, stomach/intestine, and airway, and results in hardness of the organ. The second one takes place mainly in an organ that has low potential of regeneration of cells such as skeleton muscle, and results in atrophy of the organ. By deforming the structure of tissues/organs, fibrosis reduces gradually the organ functionality and leads to development of diseases. The most known diseases associated with fibrosis are essential arterial hypertension (arteriosclerosis), senile chronic bronchitis/emphysema, and benign prostatic hyperplasia.

Arteriosclerosis is the main pathology in essential arterial hypertension. Arteriosclerosis is characterized by intima fibrosis. The stiffness of part of arterial wall caused by intima fibrosis increases the resistance of the local wall against blood flow. The consequences are: **A**. the local arterial wall becomes fragile under the load of blood flow; **B**. the blood supply to local tissue is reduced; and **C**. the load of blood flow to the heart is increased. Thus, progressive arteriosclerosis increases gradually the risk of cerebral bleeding, cerebral infarction, aneurysm, and heart failure. Similarly, fibrosis of airway wall is a characteristic change in senile chronic bronchitis/emphysema. Hardness of part of airway wall increases the resistance of the wall against airflow. Deposition of excessive air in the lung leads to emphysema and disruption of alveolus. Benign prostatic hyperplasia is pathologically characterized by loss of myofibers, deposition of collagen fibers in muscular tissues, and expansion of gland tubes. The enlarged and stiffened prostate will finally disturb the functions of its neighbor organs. With age, many of our organs become atrophic because of tissue fibrosis, including skin, skeleton muscle, and gland organs. Diseases will occur when an atrophied organ fails of functions. For example, senile type II diabetes is found to be a result of atrophy of the islet. B-cells in the islet are the cells producing insulin. Some old people have reduced number of B-cells but increased amount of collagen fibers in their islets.

Tissue fibrosis is a typical aging change; however traditional aging theories, including gene-controlling theory (Bell, 2012), free-radical theory (Harman, 1956), cell senescence/telomere theory (Hayflick, 1965), and damage (fault)-accumulation theory (Kirkwood, 2005) cannot interpret this phenomenon. Therefore, these theories are untenable. Medical studies have shown that tissue fibrosis is a result of repair of tissue with collagen fibers. In fact, a repair with collagen fibers is a manner of "Misrepair". Thus, our Misrepair-accumulation theory can explain exactly the phenomenon of fibrosis. In our view, fibrosis is a result of accumulation of Misrepairs of tissue with collagen fibers (Wang, 2009).

## II.    A novel aging theory: Misrepair-accumulation theory

In analyzing aging changes, we recognize that incorrect repair of a tissue/organ in an organism is a permanent change that contributes to the aging of the organism. Scar formation and DNA Misrepair are examples of incorrect repair. On this basis, we proposed a novel aging theory, the Misrepair-accumulation theory (Wang, 2009). The concept of Misrepair in this theory is generalized and defined as *incorrect reconstruction of an injured living structure.* This new concept of Misrepair is applicable to all types of living structures that are repairable, including DNAs, cells, and tissues. Structural injuries of a tissue/organ caused by external or internal damage are unavoidable. For a large injury, a Misrepair with altered materials and in altered remodeling is essential for closing the structural defect and for the surviving of the organism. Without Misrepairs, an individual could not survive to the age of reproduction; thus Misrepairs mechanism is essential for the survival of a species.

However, a Misrepair results in alteration of structure and reduction of functionality of a cell or a tissue, and the structure-alteration made by Misrepair is irreversible. Misrepairs are unavoidable and irremovable; thus they accumulate and deform gradually the structure of a molecule, a cell, or a tissue, appearing as aging of it. Aging is a process of accumulation of Misrepairs of a structure. Misrepair of a tissue leads to increased damage-sensitivity and reduced repair-efficiency of the tissue. As a consequence, new Misrepairs have a tendency to occur to the misrepaired part of the tissue. Accumulation of Misrepairs is therefore focalized and self-accelerating (Wang-Michelitsch, 2015). Aging takes place on the levels of molecules, cells, and tissues, respectively. Aging of a tissue appears as disorganization of cells/extracellular matrixes (ECMs). Aging of a cell appears as a gradual deformation of the cell and a change of the distribution of organelles and chromosomes in the cell. Aging of a DNA appears as accumulation of DNA mutations and gradual alteration of DNA sequence. However, aging of a multi-cellular organism takes place essentially on tissue level. An irreversible change of the spatial relationship between cells/ECMs in a tissue is *essential and sufficient* for causing a decline of organ functionality. In summary, aging of an organism is a result of accumulation of Misrepairs on tissue level.

## III. Tissue fibrosis: a result of accumulation of Misrepairs of tissue with collagen fibers

Chronic inflammatory diseases all end up with fibrosis of organs. For example, chronic hepatitis B results in liver cirrhosis and rheumatoid arthritis leads to fibrosis of synovial tissue. Fibrosis is a result of repeated healing of an injured tissue. Healing with collagen fibers is a way of repair; and the collagen fibers are used for replacing dead cells or ECMs. Such Misrepairs can take place in several situations, including: **A**. an injury with death of un-regenerable cells/ECMs, such as skeleton muscular cells, and basement membrane; **B**. an injury with death of a large number of cells/ECMs, such as a tissue injury by a severe accident and that in chronic hepatitis B; **C**. repeated injuries, such as the repeated disruptions of elastic fibers and myofibers in arterial wall or airway wall due to the rhythmic deformations of these organs; and **D**. an injury with deposition of un-degradable substances, such as an injury of endothelium with infusion of lipids into endothelium (Table 1). In these four situations, a Misrepair with collagen fibers for replacing the dead cells/ECMs is essential for closing the structural defect in a permitted time. If the "structural gap" is not sealed in time, failure of the structure and death of the organism will take place.

### 3.1 Collagen fibers in replacing dead cells and broken basement membrane

Some types of cells including skeleton muscular cells and neuron cells have no potential of regeneration. Death of these cells will promote Misrepair. Namely, the structural defect in skeleton tissue or neural tissue left by dead cells need to be closed by other types of cells or ECMs (Table 1). Except epithelial cells, hepatocytes, endothelial cells, and blood cells, most types of cells in an animal have low potential of regeneration. For the tissues that can reproduce cells, a small injury with occasional death of cells can be completely restored by regeneration of cells. However, in the case of a severe injury, a "big hole" of a tissue caused by death of a large number of cells/ECMs cannot be rapidly refilled by regenerated cells, even in the tissues that have great potential of regeneration. Basement membrane is a dense membrane separating the epithelium and the derma in skin and in mucosa. Basement membrane has a complex structure constructed with several types of ECM molecules. This complex structure is thus difficult to be quickly remodeled when it is broken. Scar formation is often promoted by an injury of basement membrane in a deep skin wound (Table 1). Taken together, for those tissue injuries with death of un-regenerable cells, death of a great number of cells/ECMs, or broken basement membrane, the repair has to be achieved by Misrepair. The Misrepair is made by proliferation of fibroblasts and production of collagen fibers, and this is for eliminating the tissue defect and re-linking the broken tissue.

**Table 1. Misrepairs of tissues with collagen fibers for replacing dead cells/ECMs**

| Injuries | Misrepairs | Consequences |
|---|---|---|
| **Death of un-regenerable cells/ECMs** | **Closing tissue defect by collagen fibers** | **Tissue fibrosis** |
| e.g. Death of skeleton muscular cells | | ➢ Atrophy of skeleton muscle |
| Injury of basement membrane | | ➢ Scar formation |
| **Death of a large number of cells/ECMs** | **Closing tissue defect by collagen fibers** | **Tissue fibrosis** |
| e.g. Chronic hepatitis B | | ➢ Liver cirrhosis |
| **Repeated injuries of elastic fibers or myofibers** | **Replacing the broken elastic fibers/myofibers by collagen fibers** | **Tissue fibrosis** |
| e.g. In skin | | ➢ Wrinkle formation |
| In arterial wall | | ➢ Arteriosclerosis |
| In airway wall | | ➢ Fibrosis of airway wall |
| In prostate | | ➢ Prostatic hyperplasia |
| **Deposition of un-degradable substances** | **Isolating the dead substances by collagen fibers** | **Degeneration changes with tissue fibrosis** |
| e.g. Injury of endothelium and infusion of lipids into endothelium | | ➢ Atherosclerotic plaques |

## 3.2 Collagen fibers in replacing disrupted elastic fibers and myofibers

Some organs, including the heart, artery, stomach/intestine, the lung, and the prostate, work in a manner of rhythmic deformations of their "walls". Smooth muscular cells in the walls of these organs, called myofibers, can actively regulate the rhythmic deformations of organs through periodical contracting/dilating. Myofibers can undergo enlarging when being loaded. Myofibers have certain degree of potential of regeneration, and they can proliferate when some are disrupted. In these organs, the connective tissues in surrounding of muscular tissues can passively deform due to the elasticity of elastic fibers and the deformability of the organization of collagen fibers. The elasticity of an elastic fiber comes from its special structure. The cross-linking of elastin subunits in a special manner makes the chain of subunits, an elastic fiber, changeable in length when compressed/extended (Figure 1A).

Elastic fibers are the main component of the elastic membrane in arterial wall, airway wall, and gastrointestinal wall. Differently, a single collagen fiber is thick and inelastic, with strong resistance against external dragging. Collagen fibers are responsible for maintaining the basic shape of a tissue/organ (Figure 1B). In a connective tissue, collagen fibers are constructed in a network-like organization, and such a structure gives a tissue/organ a certain degree of "deformability" (Figure 1C). For example, articular cartilage can deform when under a load due to the network-like organization of type II collagen fibers in the cartilage. The deformability of cartilage is important in relaxing the pressure between two bones in a joint during the movements of the body.

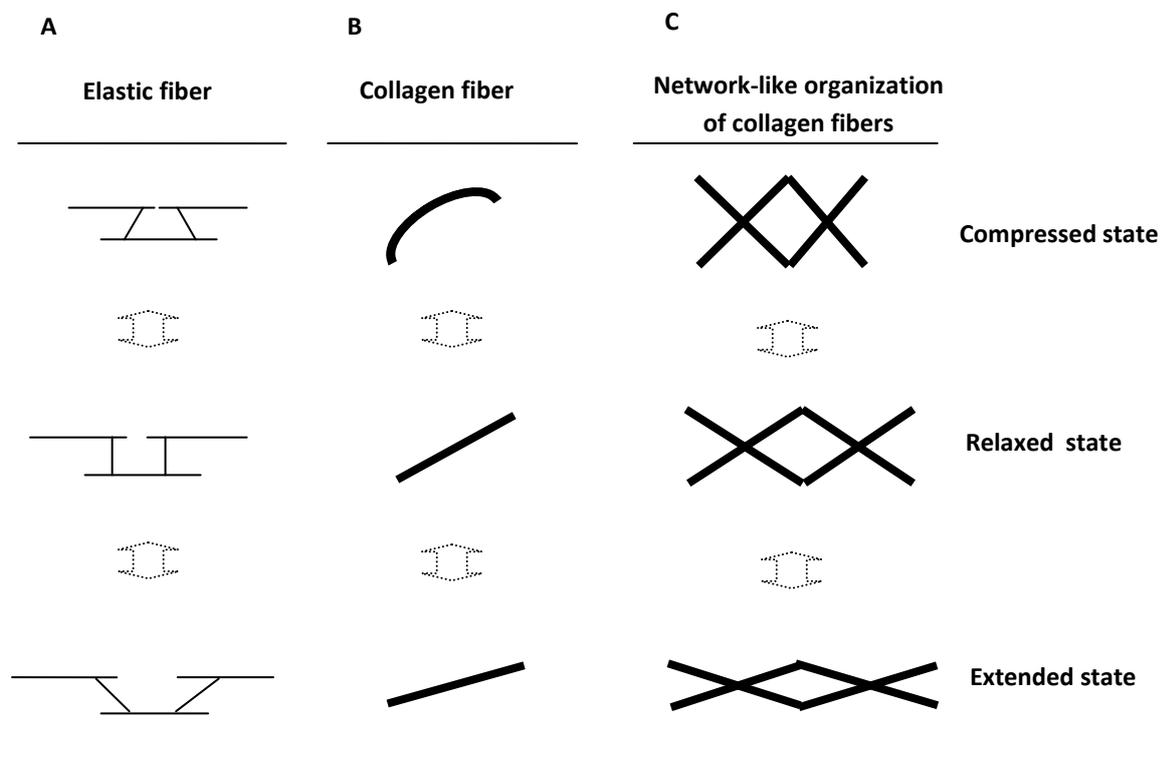

**Figure 1. Elastic structure of an elastic fiber and deformable organization of collagen fibers**

An elastic fiber has an extendable structure. Cross-linking of elastin subunits builds up an elastic fiber. A special way of cross-linking of the subunits makes the fiber changeable in length when compressed or extended (**A**). A collagen fiber is thick and inelastic, with strong resistance against external dragging. Collagen fibers are responsible for maintaining the basic shape of a tissue/organ (**B**). In a connective tissue, collagen fibers are constructed in a network-like organization, and such a organization gives a tissue/organ a certain degree of "deformability" (**C**).

In elastic organs, the repeated organ deformations are the main intrinsic source of repeated injuries of tissues. Repeated injuries, even if very small, are often difficult to be fully repaired, because the repair process is interrupted by repeatedly arrived injuries. The elastic structure of elastic fibers and the network-like organization of collagen fibers are complex structures.

These structures cannot be fully remodeled in a permitted time when severely injured or repeatedly injured. Thus, for closing quickly the tissue defect left by broken elastic fibers or myofibers, collagen fibers have to be used in emergency. The repairing collagen fibers are used for re-linking the tissue, thus they are not essentially built in a network-like organization. Like that in a scar, deposition of repairing collagen fibers reduces the deformability of the local tissue. New collagen fibers are produced by the fibroblasts or the myofibers in the tissue.

### 3.3 Collagen fibers in isolating un-degradable substances in a tissue

In a tissue, death of cells/ECMs often results in deposition of dead substances. These substances have lost functionality and need to be removed. Normally dead substances are captured and digested by local monocytes and finally cleared out by the liver-kidney clearing system. However, when dead substances are over-produced as a consequence of death of many cells in a severe injury, they cannot be removed completely. The remained dead substance, if not treated, could cause failure of the tissue, because it makes a "structural gap" in the tissue. A compromising solution for maintaining tissue functions is to isolate the un-degradable substances by a capsule. Collagen fibers, as the native materials for constructing the structure of tissues/organs and with strong physical and chemical resistance, are ideal for this purpose. A capsule made of collagen fibers is not only for isolating the dead substances but also for rebuilding a communicating pathway for the neighbor cells/ECMs. For example, Development of an atherosclerotic plaque in atherosclerosis is triggered by an endothelial injury. However, because of infusion and deposition of lipids beneath endothelial cells, the repair has to be achieved by altered remodeling of the local endothelium. A capsule made of smooth muscular cells, collagen fibers, and basement membrane is built up for isolating the lipids and for reconstructing an anchoring structure for the endothelial cells.

### 3.4 Accumulation of repairing collagen fibers: focalized and self-accelerating

Observable tissue fibrosis is a result of accumulation of repairing collagen fibers in a tissue for many years. Reduced efficiency on adaptive response and on repair due to deposition of collagen proteins makes part of a tissue have increased damage-sensitivity and increased risk for Misrepairs. For example, when some elastic fibers in part of arterial wall are broken and replaced by collagen fibers, this part of wall will have increased risk for injuries and for Misrepairs. Thus, accumulation of Misrepairs with collagen fibers is not random but focalized. Deposition of repairing collagen fibers in a tissue is not only focalized but also self-accelerating. Self-acceleration means that fibrosis of a tissue cannot stop once it starts and the rate of fibrosis will increase with time. This explains why fibrosis-related diseases are all progressive. Figure 2 shows the process of the focalized accumulation of repairing collagen fibers in arterial wall.

The layer of elastic membrane in arterial wall is in a shape of a tube, as that in Figure 2. The cross section of this membrane is in a shape of a circle. Rhythmic deformations of arterial wall and the loading of blood flow to the wall may result in repeated injuries of elastic fibers. When an elastic fiber in the circle is broken and replaced by a collagen fiber, this part of circle will have reduced elasticity and increased resistance against blood flow. The neighbor elastic fibers next to this repairing collagen fiber will have increased load from blood flow. These neighbor elastic fibers thus have increased risk to be injured and to be replaced by collagen fibers. In this way, more and more elastic fibers in the neighborhood will be broken and replaced by collagen fibers with time. The remodeled area of the elastic circle will be enlarged gradually. In the same time, the remodeled area of the elastic membrane is also enlarged along the wall (in the direction of blood flow). Finally, the focalized accumulation of repairing collagen fibers results in formation of a spot of hyaline degeneration in the tissue. With time, a spot of hyaline degeneration will grow and the number of spots will increase. Older spots are always bigger than younger ones, thus they are in an inhomogeneous distribution.

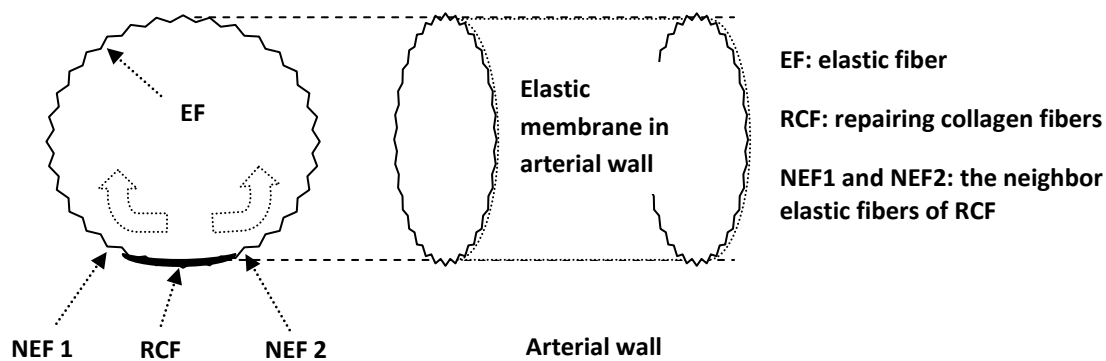

**Figure 2. Focalized accumulation of repairing collagen fibers in arterial wall**

The layer of elastic membrane in arterial wall is in a shape of a tube, and the cross section of this membrane is in a shape of a circle. Rhythmic deformations of arterial wall and the loading of blood flow to the wall may result in repeated injuries of elastic fibers. When an elastic fiber (**EF**) in the circle is broken and replaced by a collagen fiber (**RCF**), this part of circle will have reduced elasticity and increased resistance against blood flow. The neighbor elastic fibers (**NEF1** and **NEF2**) next to this repairing collagen fiber will have increased load from blood flow. These neighbor elastic fibers thus have increased risk to be injured and to be replaced by collagen fibers. In this way, more and more elastic fibers in the neighborhood will be broken and replaced by collagen fibers with time. The remodeled part of the elastic circle will be enlarged gradually (thick arrows). In the same time, the remodeled area of the elastic membrane is also enlarged along the wall. The focalized accumulation of repairing collagen fibers results in formation of a spot of hyaline degeneration in the wall.

## IV. Interpretations of fibrosis-related aging changes

Accumulation of repairing collagen fibers disorganizes gradually the structure of a tissue and reduces the functionality of an organ. Deposition of collagen fibers is a cause not only for diseases but also for some typical aging changes, including wrinkle formation and senile hair-loss/hair-whitening. In this part, we will discuss briefly how these diseases and aging changes develop by fibrosis.

## 4.1 Arteriosclerosis develops as a result of accumulation of collagen fibers in arterial wall

Arteriosclerosis occurs mainly to small arteries. Characteristic changes of arteriosclerosis in pathology are: deposition of collagen fibers in intima-media layer, proliferation of myofibers, and compensatory enlargement of myofibers in arterial wall. The deposed collagen fibers in arterial wall are the materials that are used in repair for replacing broken elastic fibers and dead myofibers. Although death of myofibers can promote proliferation of myofibers, the potential of regeneration of myofibers is low. Dead myofibers are more often replaced by collagen fibers. Deposition of collagen fibers makes the local part of arterial wall have reduced elasticity and increased resistance against blood flow. Increased resistance will induce the enlargement of local myofibers for functional compensation. Finally, this part of arterial wall becomes not only stiff but also thick, and the local arterial lumen becomes narrow. In return, increased resistance to blood flow again makes the local wall have increased risk for injuries and for deposition of collagen fibers. Repeated contractions/dilatations of arteries make the process of accumulation of collagen fibers in arterial wall continue without stop. This explains why the diseases of arteriosclerosis and essential hypertension are progressive with age.

## 4.2 Benign prostatic hyperplasia develops as a consequence of muscular fibrosis in prostate

Prostate is a gland organ for producing and excreting prostatic fluid. The muscular tissue in prostate provides the pushing force for excreting prostatic fluid during ejaculation. Unfortunately, prostate begins expanding after age 50 in a man, and often develops into benign prostatic hyperplasia. Pathology of prostatic hyperplasia is characterized by expansion of gland tubes, loss of myofibers, and deposition of collagen fibers in the muscular tissue. Movements of contraction/dilation of myofibers produce the force for pushing out prostatic fluid; however repeated deformations of myofibers may cause injuries of myofibers. Myofibers have low potential of regeneration, and the broken myofibers often need to be replaced by collagen fibers in repair. Loss of myofibers and deposition of collagen fibers reduce the force of the muscular tissue for excreting fluid. As a consequence, part of the fluid may depose in gland tubes after each time of ejaculation. Deposition of fluid and stiffness of the muscular tissue increase in return the load of fluid to myofibers. More and more myofibers will be broken during excretions and be replaced then by collagen fibers (Figure 3). By this vicious circle, fibrosis of the muscular tissue and expansion of gland tubes continue, and the prostate becomes larger and stiffer gradually. In summary, benign prostatic hyperplasia is a result of muscular fibrosis and the subsequent deposition of prostatic fluid in gland tubes.

*email : thomasjicun@gmail.com

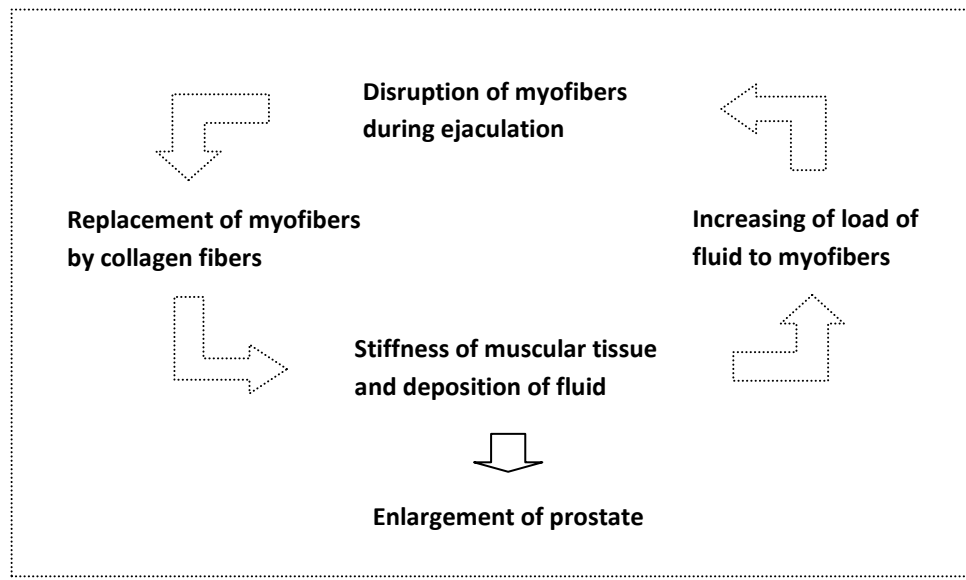

**Figure 3. Prostatic hyperplasia is a consequence of muscular fibrosis in prostate**

Movements of contraction/dilation of myofibers produce the force for pushing out prostatic fluid; however repeated deformations of myofibers during ejaculation may cause injuries of myofibers. Myofibers have low potential of regeneration, and the broken myofibers often need to be replaced by collagen fibers in repair. Loss of myofibers and deposition of collagen fibers reduce the force of the muscular tissue for excreting fluid. As a consequence, part of the fluid may depose in gland tubes after each time of ejaculation. Deposition of fluid and stiffness of the muscular tissue increase in return the load of fluid to myofibers. More and more myofibers will be injured during excretions and be replaced by collagen fibers. By this vicious circle, fibrosis of the muscular tissue and expansion of gland tubes continue, thus the prostate becomes larger and stiffer gradually. .

### 4.3 Wrinkle formation is a result of accumulation of collagen fibers of different lengths

Wrinkle formation is a typical sign of aging in human being. The number of wrinkles and the depth of a wrinkle will increase with age. In our view, wrinkle formation is a result of repair of the deeper layer of skin (derma). Our skin can be extended and compressed due to the elasticity of elastic fibers and the deformable organization of collagen fibers in dermal layer. When we smile, part of the skin on the face is extended and part is compressed. An elastic fiber can be injured when the skin is too much extended. When the injury is small and occasional, complete remodeling of the elastic fiber can take place. However, when injuries occur repeatedly, full repair of the complex structure of an elastic fiber cannot be achieved in a permitted time. For maintaining the structural integrity of skin, a "simpler" structure has to be rebuilt quickly by inelastic collagen fibers for re-linking the tissue. Otherwise the skin will be completely broken by further movements of extension/compression. However, since the repairing collagen fibers are not in a deformable organization, they can be again broken when the skin is too much compressed. The broken collagen fibers will be further replaced by new collagen fibers.

However, in different situations, the collagen fibers used for repair may have different lengths, and this is the origin for the formation of winkles. For example, when an elastic fiber is broken in an extended state, it will be replaced by a "long" collagen fiber, which is longer than that in relaxed state. Accumulation of "longer" collagen fibers will make the local skin larger and stiffer. The skin becomes flabby or prolapsing locally. When a "long" collagen fiber is broken in a compressed state, it will be replaced by a "shorter" collagen fiber. What is interesting is that a "shorter" collagen fiber will restrict the extension of the "longer" fibers. After deposition of the "shorter" fibers, the "longer" fibers have to rest permanently in a folding state, leading to the development of a wrinkle. Further accumulation of shorter and shorter collagen fibers will make the folding of longer fibers deeper and deeper. Thus, wrinkles will become deeper and deeper with time. In summary, development of a wrinkle is a result of accumulation of repairing collagen fibers of different lengths in the skin.

### 4.4 Senile hair-loss and hair-whitening are consequences of dermal fibrosis

Hairs are non-living structures that are composed of degenerated epithelial cells, termed keratinocytes. Growing of a hair is a result of cell division of the epithelial cells in the hair matrix in a follicle. The color of hairs comes from the melanin pigments in hair keratinocytes, and the pigments are produced and secreted by the melanocytes in hair matrixes. Hair matrixes obtain nutrition from the papillas in derma. Genetic defects and dysfunctions of hormones may cause early and quick loss of hairs and hair-whitening. However, since age 40, all of us will have problems of progressive hair-loss and hair-whitening, which we call "senile hair-loss" and "senile hair-whitening". Actually, our hairs do not grow all the time. For a moment, some hair matrixes are in growing state and some have stopped of growing permanently. The rests are in a pausing state and they can be reactivated into growing state. The period of time when we have hair-loss is the moment when the number of hair matrixes in growing state is less than that in pausing state and in stopping state. Pathologically, stop of growth of a hair is a result of death of all the epithelial cells in its hair matrix. Whitening of a hair is a result of death of the melanocytes in its hair matrix. Thus, permanent hair-loss and hair-whitening are results of increase of dead matrixes.

There are mainly two causes for death of a hair matrix: **A**. direct injury by external damage, such as radiation, bacterial infection, or dragging of hairs; and **B**. long-term insufficiency of nutrition or blood supply of the hair matrix. The hair-loss that is caused by external damage is local, and it will stop when the damaging element is withdrawn. Hair-loss in old people can be a result of both of direct injuries and insufficient nutrition to hair matrixes. However, a progressive hair-loss with age can only be a result of long-term insufficiency of nutrition of hair matrixes. A main cause for the insufficiency of nutrition of a tissue is the deficiency of local blood circulation. With age, derma becomes fibrotic due to accumulation of collagen fibers. The blood supply in derma will be affected by the stiffness of derma. All the papillas and follicles in the fibrotic part of derma will suffer from the shortage of blood supply. Thus, progressive dermal fibrosis is a cause for senile hair-loss.

Hair-whitening can have the same causes as that for hair-loss. However, loss of hair-color may begin earlier in age than loss of hair. The ratio of melanocytes to epithelial cells in a hair matrix is about 1/10~40. Death of all melanocytes can take place earlier than death of all epithelial cells in a hair matrix when deprived of blood supply. In summary, by affecting the blood supply to hair matrixes, dermal fibrosis is associated with the permanent and progressive hair-loss and hair-whitening in old people.

### 4.4 Senile atrophy of the brain is due to "fibrosis" of nerve tissue?

Loss of neuron cells is significant in aged brains. It is known that dead neuron cells will be replaced *in situ* by glial cells in repair. Replacement of dead neuron cells by glial cells is a manner of Misrepair. Such a Misrepair is essential for maintaining the integrity of the complex network of nerve fibers in neural tissue. Neuron cells cannot be regenerated, thus Misrepair with glial cells is the main way of repair of nerve system. However, accumulation of replacing glial cells may result in a kind of "fibrosis" of neural tissue. With time, the reduction of number of neuron cells and the increase of number of glial cells will result in atrophy of the brain. Some senile diseases in neural system may be results of atrophy of the brain. For example, Parkinson's syndrome is a disease caused by insufficiency of production of dopamine in the brain. However, the insufficiency of dopamine is often a result of loss of neuron cells in basal ganglia and substantial nigra due to atrophy of the brain. In senile neural deaf, loss of hearing is often a result of loss of the nerve cells that are responsible for transmitting neural signals from the ears to the brain. Thus, "fibrosis" may also take place in the brain, but rather made by glial cells. This "fibrosis" is associated with the degeneration of brain tissue and atrophy of the brain.

### V.  Conclusions

Tissue fibrosis is a typical aging change; however traditional aging theories cannot interpret this phenomenon. Exceptionally, our Misrepair-accumulation theory can interpret well the cause and the unavoidability of fibrosis in an organism. Tissue fibrosis is a result of accumulation of Misrepairs of tissue with collagen fibers. The collagen fibers are used for replacing dead cells and broken ECMs, including basement membrane, elastic fibers, and myofibers.  The progressive process of tissue fibrosis with age in our body manifest the existence of Misrepair mechanism, including: **A**. a process of Misrepair exists; **B**. Misrepairs are unavoidable; and **C**. Misrepairs accumulate.  Thus, the phenomenon of fibrosis is a powerful proof for the central role of Misrepair in aging. As a result of accumulation of Misrepairs, tissue fibrosis is focalized, inhomogeneous, and self-accelerating.  This explains why fibrosis-associated diseases are all progressive with time.